\documentclass[aps,a4paper, floats,nofootinbib,amssymb,superscriptaddress,twocolumn]{revtex4}

\usepackage[sort&compress]{natbib}
\usepackage{epsf,epsfig}
\usepackage{amsmath}
\usepackage{hyperref}
\usepackage{subfigure}
\usepackage{graphicx}
\usepackage{color}
\usepackage{mathrsfs}
\usepackage{enumitem}

\newcommand{\be}{\begin{equation}}
\newcommand{\ee}{\end{equation}}
\newcommand{\bea}{\begin{eqnarray}}
\newcommand{\eea}{\end{eqnarray}}

\newcommand{\ba}{\begin{array}}
\newcommand{\ea}{\end{array}}
\newcommand{\bi}{\begin{itemize}}
\newcommand{\ei}{\end{itemize}}

\newcommand{\wpwm}{W^{+}W^{-}}

\newcommand{\myspace}{\vspace{0.2cm} \\ }

\newcommand{\ih}{-\frac{i}{2}}

\makeatletter
\begin{document}

%-----------------------------------
% Preprint numbers
%-----------------------------------

%-----------------------------------
% Title
%-----------------------------------
%\begin{frontmatter}

\title{\vspace*{1.in}
\Large{Accommodating the H$(650)$ in the HEFT}
\vspace*{0.5cm}
}

%-----------------------------------
% Authors
%-----------------------------------

%\author[label4]{\large I\~nigo  Asi\'ain}
%\emailauthor{iasiain@icc.ub.edu}{I\~nigo  Asi\'ain}
%\author[label4]{\large Dom\`enec Espriu}
%\emailauthor{espriu@icc.ub.edu}{Dom\`enec Espriu}
%\author[label4]{\large Federico Mescia}
%\emailauthor{mescia@ub.edu}{Federico Mescia}
%\address[label4]{Departament de F\'isica Qu\`antica i Astrof\'isica, Institut de Ci\`encies del Cosmos (ICCUB), \\  Universitat de Barcelona, Mart\'i i Franqu\`es 1, E-08028 Barcelona, Spain}
%
\author{I\~nigo  Asi\'ain}
\email{iasiain@icc.ub.edu}
\affiliation{Departament de F\'isica Qu\`antica i Astrof\'isica\,,Institut de Ci\`encies del Cosmos (ICCUB), \\Universitat de Barcelona, Mart\'i Franqu\`es 1, 08028 Barcelona, Spain}
\author{Dom\`enec Espriu}\email{espriu@icc.ub.edu}
\affiliation{Departament de F\'isica Qu\`antica i Astrof\'isica\,,Institut de Ci\`encies del Cosmos (ICCUB), \\Universitat de Barcelona, Mart\'i Franqu\`es 1, 08028 Barcelona, Spain}
\author{Federico Mescia}\email{mescia@ub.edu}\affiliation{Departament de F\'isica Qu\`antica i Astrof\'isica\,,Institut de Ci\`encies del Cosmos (ICCUB), \\Universitat de Barcelona, Mart\'i Franqu\`es 1, 08028 Barcelona, Spain}
%\thispagestyle{empty}

%-----------------------------------
% Abstract
%-----------------------------------
\begin{abstract}
Loss of unitarity in an effective field theory is often cured by the appearance of dynamical resonances, revealing the
presence of new degrees of freedom. These resonances may manifest themselves when suitable unitarization techniques
are implemented in the effective theory, which in the scalar-isoscalar channel require making use of the coupled-channel formalism. Conversely, experimental detection of a resonance may provide interesting information on the couplings and constants
of the relevant effective theory. By applying the systematical procedure developed in previous works,
we will attempt to accommodate a  possible scalar resonance with mass around $650$ GeV for which there is preliminary
evidence at the LHC in the vector boson fusion channel. The results are interesting: the resonance can be accommodated
within the experimentally allowed range of next-to-leading order coefficients in the HEFT but in a rather non-trivial manner.
Interestingly, its width and production cross  section turn out to agree with the tentative experimental results.  
\end{abstract}

%\end{frontmatter}

\maketitle

%\thispagestyle{empty}
 %\setcounter{page}{0}

%%%%%%%%%%%%%%%%%%%%%%%%%%%%%%%%%%%%%%%%%%%%%%%%%%%%%%%%%
\section{Introduction}\label{sec: introduction}
The presence of new particles at the LHC would be a clear indication of the existence of new physics beyond the Standard Model (SM).
The masses of these yet undiscovered states could suggest a scale of new physics, to be explored, and their production channel
would provide a guidance for future-experiment proposals. However, after the discovery in 2012 of a Higgs-like particle, the $h(125)$,
so far compatible with the minimal SM, there have been no clear signals of new findings regarding this matter.\myspace
There is an obvious interest in the appearance of (relatively) light scalar companions of the $h(125)$ as they may arise in
composite Higgs models (see, e.g., Ref.~\cite{Dobado:2019fxe} and the references therein) as (pseudo) Goldstone-like spare states following the spontaneous symmetry breaking (SSB) of the
vacuum of a theory possessing a larger global symmetry group. Other models try to explain such triggering of the \textit{electroweak symmetry breaking sector} (EWSBS) with more than one scalar such as the Georgi-Machacek~\cite{Georgi:1985nv} and Chanowitz-Golden~\cite{Chanowitz:1985ug} and two-Higgs-doublet (2HDM)~\cite{Gunion:2002zf} models. However, up to now, experimental evidence in this direction is scarce,
with not very conclusive significance for some scalar resonances appearing in the literature such
as H$(650)$, h$^{\prime}(515),\, A(400),\, h(151),\,h(95)$~\cite{Kundu:2022bpy}.\myspace
In Refs.~\cite{Asiain:2021lch, Asiain:2023zhx} we saw in the context the Higgs Effective Field Theory (HEFT) how indeed resonant
states appear at the scale of unitarity violation of the perturbative amplitudes for certain ranges of the effective couplings.
The Inverse Amplitude Method (IAM), that derives from analytical properties of partial waves in the $s-$complex plane, is the tool
of choice employed to understand the emergence of dynamical resonances. In the case of the $IJ=00$ channel, where the $h(125)$ belongs,
the analysis is  more involved than when searching for vector resonances due to the need of making use of the coupled channel formalism.
A rather detailed exploration of possible scalar resonances and their implication for the HEFT coefficients was carried
out in Ref.~\cite{Asiain:2023zhx} but restricted to resonances heavier that 1.8 TeV. This restriction arises from the need of fulfilling
various phenomenological constraints for vector resonances \cite{Rosell:2020iub}. However, pseudo Goldstone bosons in the scalar channel may be lighter and
it is therefore appropriate to examine the possible presence of light states.\myspace
Recently, some interest has emerged on a possible signal for a Higgs-like state around 600 GeV, much below the region just mentioned.
On one hand, searches in CMS \cite{CMS:2020tkr} and ATLAS \cite{ATLAS:2021kog}, see Refs. \cite{Cea:2018tmm, Cea:2022zgs} for a combined analysis, have yielded some evidence for the production of this resonance through the clear
four-leptonic final state: H$(650)\to ZZ\to 4l$. In particular, they suggest a scalar state peaking at $\sim$ 650 GeV with a total width of
approximately 100 GeV,  with a $3.75\sigma$ significance using an integrated luminosity of 139 fb$^{-1}$. The corresponding cross section for the subprocess $pp\to ZZ+X$ is $90\pm 25$ fb. After applying sequential cuts, an ATLAS analysis for vector boson fusion \cite{ATLAS:2021kog} reduces the significance of this resonance to $2.1\sigma$ and
the cross section to $30\pm15$ fb, significantly below the inclusive one before the cuts. On the other hand, searches of leptonic decays
from $WW$ ($2l+$missing energy) enhances the production rate for this scalar to more than five times the $ZZ$ one, resulting in a
cross section of $160\pm50$ fb. This scenario of unbalanced production rates between channels will actually be reproduced in our HEFT description,
as we will see.\myspace
The question that naturally emerges is: is such a light resonance compatible with existing bounds on the low energy coefficients
  of the HEFT? This is not obvious at all, because strict bounds already exist on many of these coefficients as we will see below. These
  bounds place various such coefficients in the $10^{-4}$ range, which typically provide resonances above the TeV scale, but several other
  couplings are poorly bounded or not bounded at all.
  Can therefore the $\text{H(650)}$ be accommodated in the HEFT without violating  any existing bounds? This seems a relevant question because
  a negative answer -taking into account the generality of
  the HEFT approach- would most likely rest credibility to the experimental hints.\myspace
  In Section \ref{sec: lagrangian} we present the theoretical framework we use, namely the HEFT, with all the considerations
  we have taken in order to simplify the computation of the relevant $2\to 2$ processes at the one-loop level. We also include
  information regarding the experimental status of the couplings that define the relevant parameter space for our purposes.
  To finish this section we will succinctly comment on the systematics to build the partial waves that will be rendered unitary.
  The interested reader may find much more detailed information in Ref. \cite{Asiain:2023zhx}.\myspace
  Section \ref{sec: h650} will be devoted to the analysis of the HEFT parameter space selected by a H$(650)$-like resonance
  in $WW$ unitarized scattering.\myspace
  Some previous works have already studied models with states similar to H$(650)$~\cite{Kundu:2022bpy, Afonin:2022qkl,Georgi:1985nv,Chanowitz:1985ug,Cea:2018tmm,Cea:2022zgs,Gunion:2002zf}.
\section{Effective Lagrangian, experimental bounds and partial waves}
\label{sec: lagrangian}
In this section we will summarize our notation and, in particular, identify the low-energy constants called to play a role in the subsequent analysis.
Following previous studies in Refs.~\cite{Asiain:2021lch} and \cite{Asiain:2023zhx} regarding vector and scalar resonances  we
work in the framework of the HEFT, an $SU(2)_L\times SU(2)_R$ symmetric chiral Lagrangian with the addition of a light Higgs with mass $M_h=125$ GeV,
and under the assumption that the custodial symmetry remains exact after the spontaneous breaking of the vacuum of the theory following the
pattern $SU(2)_L\times SU(2)_R\to SU(2)_V$, therefore neglecting the soft breaking induced by $\mathcal{O}(g^{\prime})$ pieces gauging the $U(1)_{Y}$
subgroup. Consequently, we set $g^{\prime}=0$ and the purely electromagnetic effects that make the $W$ and $Z$ gauge bosons masses differ are absent:
$W$ and $Z$ transform exactly as a triplet under the custodial group (we will refer to them indistinctly as $W$) and there are no vertices
involving photons whatsoever. This simplification, useful to employ an exact weak isospin formalism, is not expected to have any significant
effect on the analysis.\myspace
The HEFT is constructed as an expansion in powers of the momentum (derivatives) and, in clear contrast to the linear case where order by order suppression is performed by explicit powers of an energy cut-off in the denominator to obey canonical dimensional analysis, the chiral order of
any operator represents the number of derivatives and/or soft mass scales $M_W$ ($\sim g$) and $M_H$ ($\sim\sqrt{\lambda}$) that it contains.
Up to $\mathcal{O}(p^4)$ we need the following pieces: 
\begin{equation}\label{eq: lag2}
\begin{split}
\!\!\!\! \mathcal{L}_2 =&-\frac{1}{2g^2}\text{Tr}\left(\hat{W}_{\mu\nu}\hat{W}^{\mu\nu}\right)-
\frac{1}{2g^{\prime 2}}\text{Tr}\left(\hat{B}_{\mu\nu}\hat{B}^{\mu\nu}\right)\\
  &+\frac{v^2}{4}\mathcal{F}(h)\text{Tr}\left(D^{\mu}U^{\dagger}D_{\mu}U\right)
+\frac{1}{2}\partial_{\mu}h\partial^{\mu}h -V(h) \vspace{0.2cm} 
\end{split}
\end{equation}
\begin{equation}\label{eq: lag4}
\begin{split}
\!\!\!\!  \mathcal{L}_4 =&-i a_3\text{Tr}\left(\hat{W}_{\mu\nu}\left[V^{\mu},V^{\nu}\right]\right)
  +a_4 \left(\text{Tr}\left(V_{\mu}V_{\nu}\right)\right)^2\\
  &+a_5 \left(\text{Tr}\left(V_{\mu}V^{\mu}\right)\right)^2
  +\frac{\delta}{v^2}\left(\partial_{\mu}h\partial^{\mu}h\right)\text{Tr}\left(D_{\mu}U^{\dagger}D^{\mu}U\right)\\
&  +\frac{\eta}{v^2}\left(\partial_{\mu}h\partial_{\nu}h\right)\text{Tr}\left(D^{\mu}U^{\dagger}D^{\nu}U\right)\\
&+\frac{\gamma}{v^4}\left(\partial_{\mu}h\partial^{\mu}h\right)^2+i\frac{\zeta}{v}\,\text{Tr}\left(\hat{W}_{\mu\nu}V^{\mu}\right)\partial^{\nu}h
\end{split}
\end{equation}
with the building blocks
\begin{equation}\label{eq: building_blocks}
\begin{split}
\!\!\!\!  &U=\exp\left(\frac{i\omega^a\sigma^a}{v}\right) ,\:\mathcal{F}(h)=1+2a\left(\frac{h}{v}\right)+b\left(\frac{h}{v}\right)^2+ \ldots , \\
  &D_{\mu}U=\partial_{\mu}U+i \hat{W}_{\mu} U, \: \hat{W}_{\mu}=g\frac{\vec{W}_{\mu}\cdot\vec{\sigma}}{2}, \: V_{\mu}=D_{\mu}U^{\dagger}U , \\
  &V(h)=\frac{1}{2}M_h^2h^2+d_3\lambda v h^3+d_4\frac{\lambda}{4}h^4+ \ldots, \\
 &\hat{W}_{\mu\nu}=\partial_{\mu}\hat{W}_{\nu}-\partial_{\nu}\hat{W}_{\mu}+i\left[\hat{W}_{\mu},\hat{W}_{\nu}\right]
\end{split}
\end{equation}
The effective Lagrangian suitable for our purposes is then
\begin{equation}\label{eq: total_lag}
\mathcal{L}=\mathcal{L}_2+\mathcal{L}_4+\mathcal{L}_{GF}+\mathcal{L}_{FP}
\end{equation}
where the last two pieces are the gauge-fixing and the associated Faddeev-Popov, respectively, that are trivial (induce no dynamics) in the
Landau gauge ($\xi=0$) with massless Goldstones that we use throughout.\myspace
The deviations from the SM are parameterized by the -often called anomalous- couplings accompanying the local operators in Eqs. (\ref{eq: lag2}) and (\ref{eq: lag4}). Any beyond-the-SM (BSM) model can be reproduced by a suitable choice of
these anomalous couplings in the HEFT. The SM corresponds to a particular choice of the couplings, 
namely $\alpha_{p^2}=1$ and $\alpha_{p^4}=0$ where $\alpha_{p^{k}}$ generically represents the full set of chiral parameters belonging
to the Lagrangian $\mathcal{L}_{p^k}$, this is, of chiral order $k$. The reverse is not true; random values for the anomalous couplings
will typically lead to inconsistencies, such as lack of causality~\cite{Adams:2006sv}, and cannot correspond to any meaningful UV completion.\myspace
We will restrict the possible values of the anomalous couplings by making use of the experimental bounds available up to date and the
hierarchy of the effects that they produce in our results. As a working hypothesis, we will assume that all $\alpha_{p^2}$ couplings take
their canonical SM values and, consequently, departures will be described by anomalous values of the $\alpha_{p^4}$.
Regarding the latter, it is easy to see to see why considering $a_3$ and $\zeta$ will have a subleading role: they are couplings
that enter at $\mathcal{O}(p^4)$, just like $a_4,\,a_5\cdots$, but they have
one derivative less [they are $\mathcal{O}(g)$ instead] which translates into one power less of momenta easing their high energy contribution
in comparison to operators with four derivatives. Thus, we will not take them into account in our analysis.\myspace
From an experimental point of view, some of the couplings
happen to be poorly restricted, or not restricted at all in the existing literature. In particular we will admit values of the couplings
in $\mathcal{L}_4$ in the range~\cite{CMS:2019uys}
\begin{equation}\label{eq: a4a5_bounds}
a_4\in (-0.0061,\,0.0063)  \hspace{0.5cm} a_5\in (-0.0094,\,0.0098) \,
\end{equation} 
that have been obtained using $13$ TeV LHC data in four leptons final states from $WW/WZ$ scattering. There is a more strict bound
(by a factor 10~\cite{Sirunyan:2019der}) for $a_5$ coming from an SMEFT analysis with $2l2j$ final states, much lesser clear channel.
However, it should be noted that in the scalar-isoscalar channel the couplings $a_4$ and $a_5$ always appear in the combination
$5a_4 +8a_5$ and therefore the error in $a_4$ amply dominates anyway.\myspace
As said, the rest of the $\alpha_{p^4}$ couplings relevant for the present discussion, namely $\delta,\,\eta$ and $\gamma$,
remain unconstrained experimentally, but taking into
account the fact that they are absent in the SM, we will allow these to have a maximum (absolute) value of $10^{-3}$. \myspace
As anticipated in the introduction the amplitudes coming from the Lagrangian (\ref{eq: total_lag}) lack  unitarity, being fastly growing
with the center of mass energy, unless all couplings are taken equal to their SM value. This fact has to be addressed if one wants to make
predictions using an effective theory that somehow keeps track of the physical UV behavior of the complete theory
from which it supposedly comes from. Hence, unitarization methods are required beyond a certain energy range. Among the various
unitarization techniques we will be using
the Inverse Amplitude Method (IAM) that has been proven to show the same qualitative results that the others and the unitarized amplitudes
match, by construction, the perturbative ones at low energies, before unitarity is manifestly lost.\myspace
The IAM is implemented in amplitudes with well-defined angular momentum and weak isospin quantum numbers, an $IJ$ basis. A way to build this amplitudes, and in particular the isoscalar-scalar ($IJ=00$) we are interested in, is presented in Ref.~\cite{Asiain:2021lch}
and it turns out to be greatly simplified in our custodial limit making use of Bose and crossing symmetries. \myspace
The result for the unitarized partial wave is
\begin{equation}\label{eq: tIAM}
\begin{split}
&t^{IAM}_{IJ}=t_{IJ}^{(2)}\cdot\left(t_{IJ}^{(2)}-t_{IJ}^{(4)}\right)^{-1}\cdot t_{IJ}^{(2)}\\
&t_{IJ}^{(n)}=\frac{1}{64\pi}\int_{-1}^{+1}d\cos\theta\,T_{I}^{(n)}(s,\cos\theta) P_J(\cos\theta)
\end{split}
\end{equation}
which at next-to-leading (NLO) precision, the formula for $t^{\cal U}$ coincides for both vector and tensor, where $t_{20}^{(n)}$ and $t_{11}^{(n)}$ are
functions of $s-$complex values, and $t_{00}^{(n)}$ that are matrices containing the coupled channels. The expressions to
relate the fixed-isospin amplitudes, $T_I$, with the amplitudes in the charged basis are gathered in Ref.~\cite{Asiain:2023zhx}.
\section{$\textbf{H(650)}$ via VBF in the HEFT}\label{sec: h650}
\begin{figure}
\centering
\includegraphics[clip,width=8.5cm,height=8.5cm]{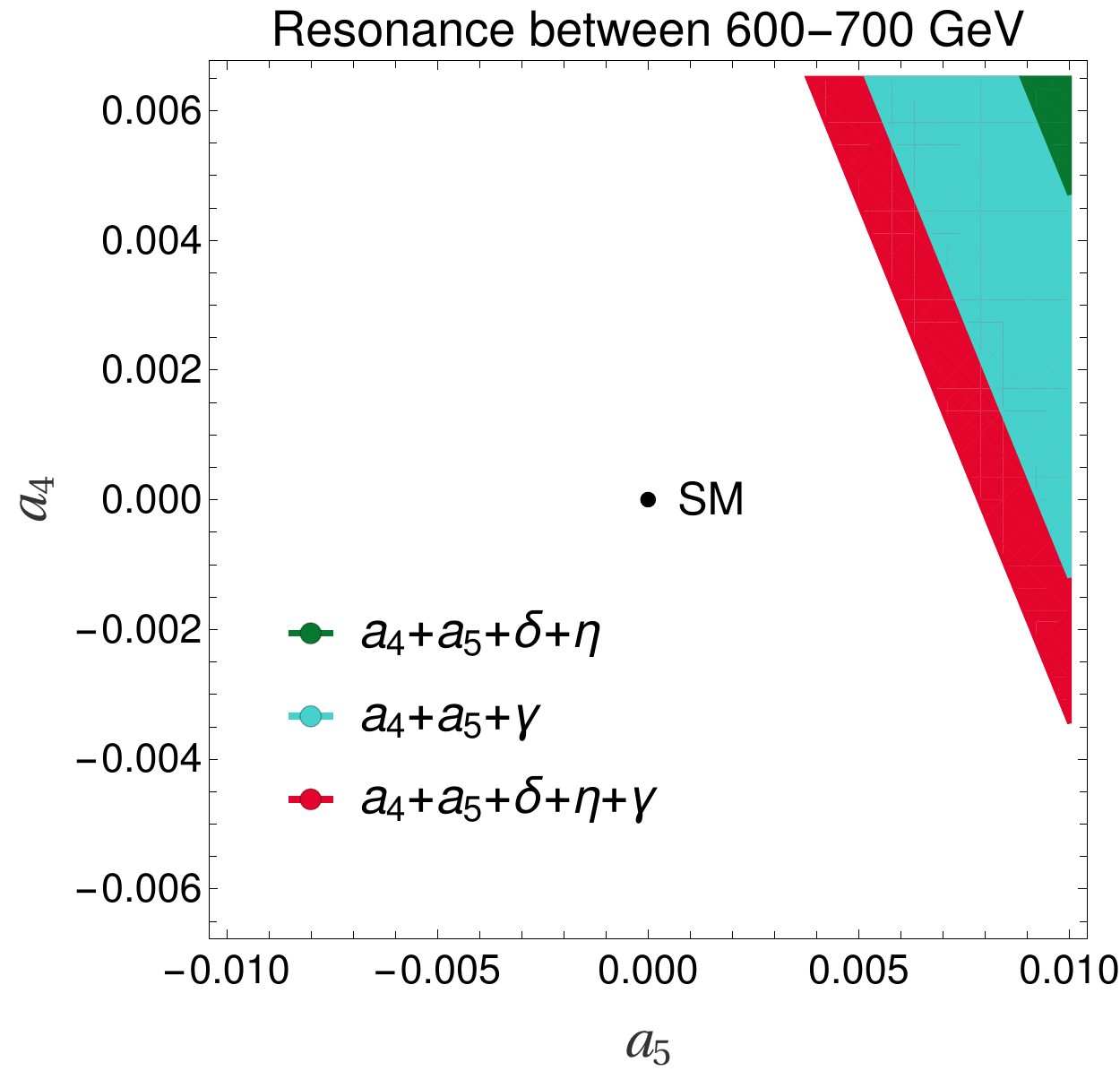}
\caption{\small{Regions in the $a_4-a_5$ plane allowed by experimental constraints where resonances between $600-700$ GeV appear when activating different NLO chiral couplings. The more parameters one activates, the less restriction in the plane to achieve the desired scalar light resonance. The LO parameters of Eq.~(\ref{eq: lag2}) are set to the corresponding SM values.}}
\label{fig: map_a4a5}
\end{figure}

\begin{figure*}
\centering
\includegraphics[clip,width=8.1cm,height=9.0cm]{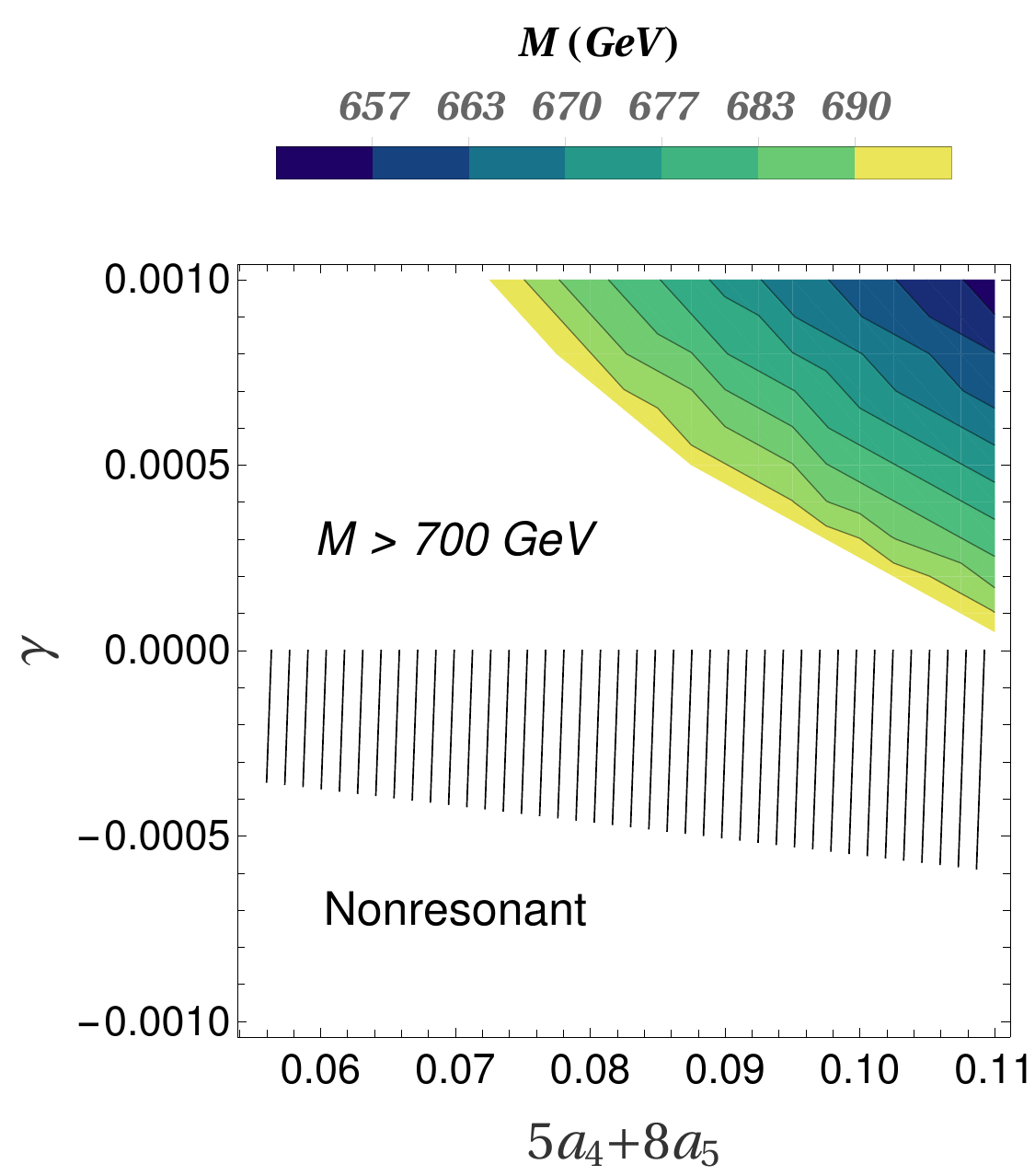}\qquad\quad
\includegraphics[clip,width=8.1cm,height=9.0cm]{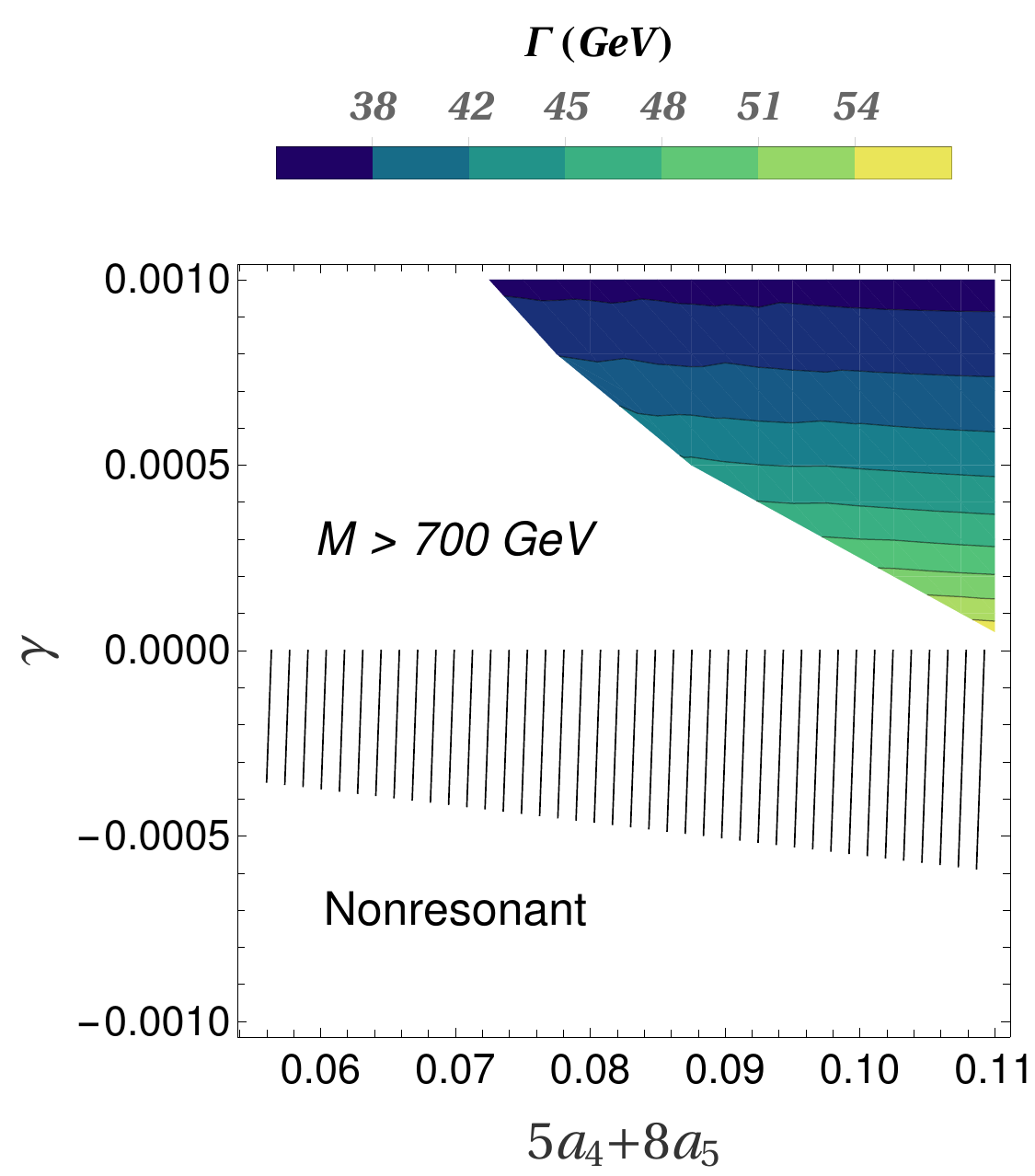}
\caption{\small{Masses (left panel) between 600 and 700 GeV and widths (right panel) of scalar resonances in the plane described by the set of lines $5a_4+8a_5=k$ and $\gamma$. All negative values of $\gamma$ are non-physical (stripped area) by the phase-shift criteria or nonresonant. Positive values of $\gamma$ below the colorful region exhibit physical resonances, too, but they are heavier than $700$ GeV so they are of no interest for this work. The right panel shows how for a fixed value of $\gamma$, the widths of the corresponding resonances are independent of the combination $5a_4+8a_5$.}}
\label{fig: map_Kgamma}
\end{figure*}

The exercise we want to do in this section is to search for a  set of $\alpha_{p^4}$ HEFT parameters that lead to the presence of a
resonance with the properties of the H$(650)$ in $WW$ scattering that are tentatively claimed in Ref.~\cite{Kundu:2022bpy}. Though the coupled channel
formalism~\cite{Asiain:2023zhx}, the elastic $WW$ channel is also coupled to both $WW\to hh$ and $hh\to hh$ at the level of unitarized
scalar waves. Experimentally, this resonance appears to have a total width of $\sim 100$ GeV so, so we focus on the production of a scalar resonance whose mass
lies within the $600-700$ GeV range.\myspace
The anomalous parameters of chiral order two are all set to their SM values: $\alpha_{p^2}=1$. This leaves us with free $a_4,\,a_5,\,\delta,\,\eta$
and $\gamma$. All of them intervene at the NLO (formally tree level) contributions for different processes: the first two to $WW$ elastic
scattering, $\gamma$ to elastic $hh$ and $\delta$ and $\eta$ to the crosses channel $WW\to hh$. However, these separated contributions mix
among themselves along the unitarization process.\myspace
Following the results in Ref.~\cite{Asiain:2023zhx}, when we set $\alpha_{p^2}=1$ there are only two physical situations for any
choice of the $\alpha_{p^4}$ chiral parameters: a nonresonant scenario with the absence of any complex pole in the unitarized amplitude, or
a resonant one with only one such pole. Whenever there are two poles, one is identified as non physical by the phase shift criteria (lies
on the first Riemann sheet in the complex $s$ plane).
Secondly, the resonances emerging are much more visible in the $WW$ channel than in the coupled ones so the $a_4-a_5$ plane is the most
sensible parameter space to represent the results. As already mentioned in a previous section, this plane is restricted by the experimental
bounds quoted in Eq.~(\ref{eq: a4a5_bounds}).\myspace
In Fig. \ref{fig: map_a4a5} we show the regions in $a_4-a_5$ parameter space where a resonance with mass between $600-700$ GeV appears using
different selection of the $\alpha_{p^4}$ chiral parameters. The different areas are obtained by activating different sets of the NLO HEFT coefficients besides $a_4$ and $a_5$, with maximum values of $|10^{-3}|$, following the explanation in the legend.
It should be clarified that all the regions overlap with each other: the red one includes the rest of the areas,
and the blue one the smaller one in green.\myspace
In fact, no bound state with the expected characteristics appears assuming non-zero values for $a_4$ and $a_5$ only. One needs the help of at least
one more anomalous coupling.\myspace
\myspace
The Fig. \ref{fig: map_a4a5} above shows regions in the main parameter space where resonances with masses in the range 600-700 GeV are allowed but it says nothing about their properties. One thing that we can indeed extract from Fig. \ref{fig: map_a4a5} is that the inclusion of $\delta$ and $\eta$ does not affect very much the results when looking for light resonances. With these, we now investigate the physical properties of the resonances using only $a_4,\,a_5$ and $\gamma$. Actually, these \textit{a priori} independent three parameters are reduced to two when studying scalar resonances since, as said before, the lines $5a_4+8a_5=k$ contain resonances with the same properties for a fixed $k$. We choose $k\in (0.055,0.11)$ so we lie within the pure scalar region, not vector nor tensor states appear, and the experimental bounds in Eq.~(\ref{eq: a4a5_bounds}) are satisfied.\myspace
In Fig. \ref{fig: map_Kgamma} we show the results. As one can observe, no scenario  with negative values of $\gamma$ exhibits physical resonances: the dashed region is forbidden by the emergence of a second pole identified as non-physical using the phase-shift criteria and below this area, a nonresonant region appears. Positive values of $\gamma$ that are not colored in Fig. \ref{fig: map_Kgamma} also lead to physical resonances heavier than $700$ GeV, though, so they are excluded from our analysis. Another interesting thing we have observed in the right panel is that for a specific value of $\gamma$ the width remains constant along the lines $5a_4+8a_5=k$ in the region of interest and they are relatively small (with respect to the masses) leading to quite stable intermediate states emerging in $WW$ scattering. This feature can be explained with the information provided in Ref.~\cite{Asiain:2023zhx}, where we observed that for large (yet natural) values of $\gamma\left(\sim 10^{-3}\right)$, the single-channel resonance, this is ignoring coupled-channels, was recovered.
\myspace
All in all, we have been able to reproduce a scalar resonance with mass around 650 GeV in the HEFT using SM values of the LO Lagrangian and deviations at the next-to-leading order in chiral counting that are within the existing experimental bounds up to date. The more relevant couplings to describe such a resonance happen to be those driving the elastic NLO processes, $a_4$, $a_5$ and $\gamma$ with subleading effects with the off-diagonal ones, $\delta$ and $\eta$, along the coupled-channel unitarization process. The widths obtained for those new states are quite small compared to their masses, 30-65 GeV.\myspace
However, up to now, the more data the experiment collects the more compatible the anomalous couplings are with the successful SM. The BSM H(650)-like resonances in this study appear close to the upper limit of the experimental bounds in Eq.~(\ref{eq: a4a5_bounds}), see Fig. \ref{fig: map_a4a5},  meaning that a possible future improvement in these bounds pointing towards consolidation of the SM values would imply their exclusion. 
\section{Comparison with literature}\label{sec: literature}
\begin{figure*}
\centering
\includegraphics[clip,width=8.1cm,height=8.1cm]{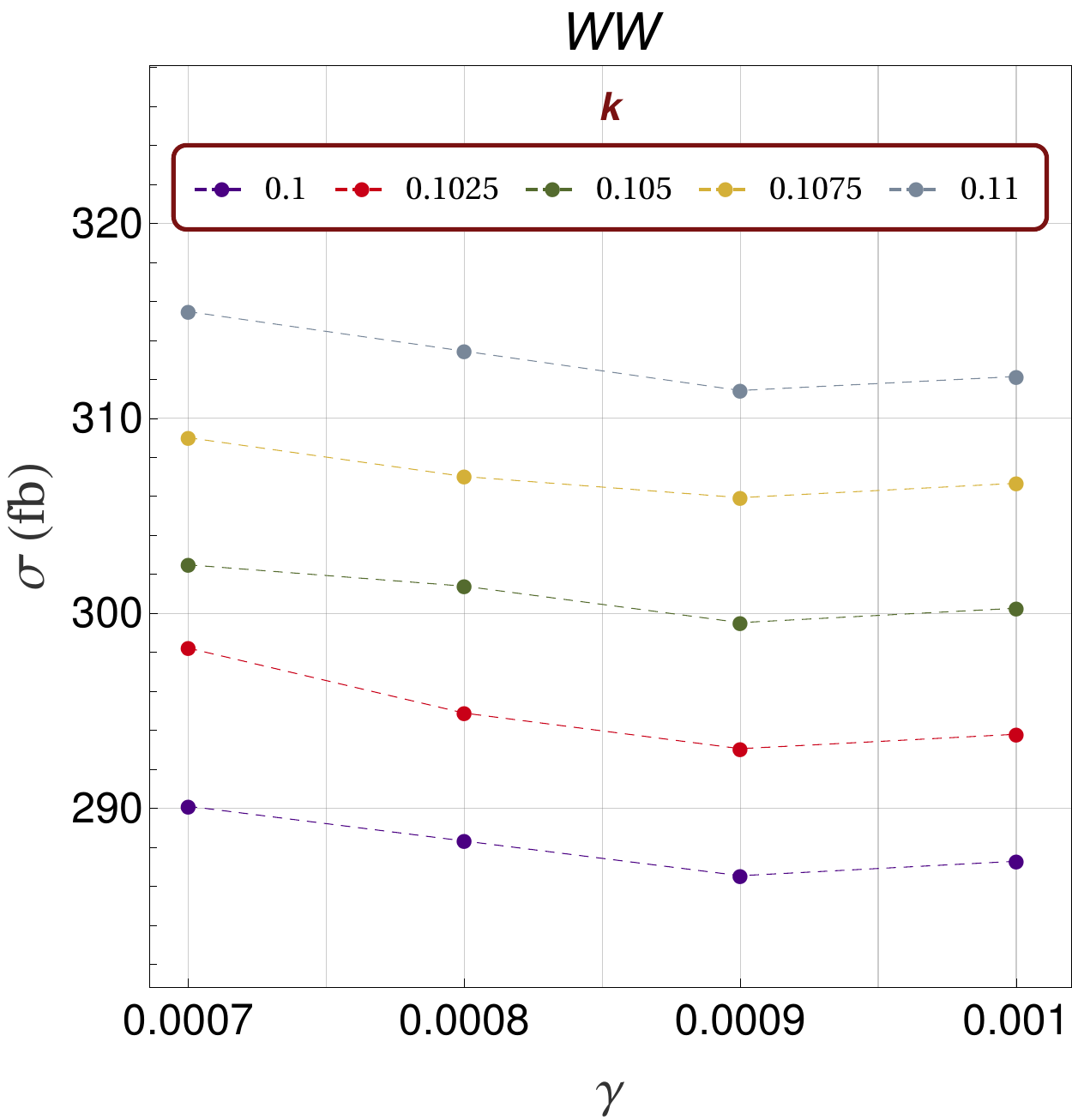}\qquad\qquad
\includegraphics[clip,width=8.1cm,height=8.1cm]{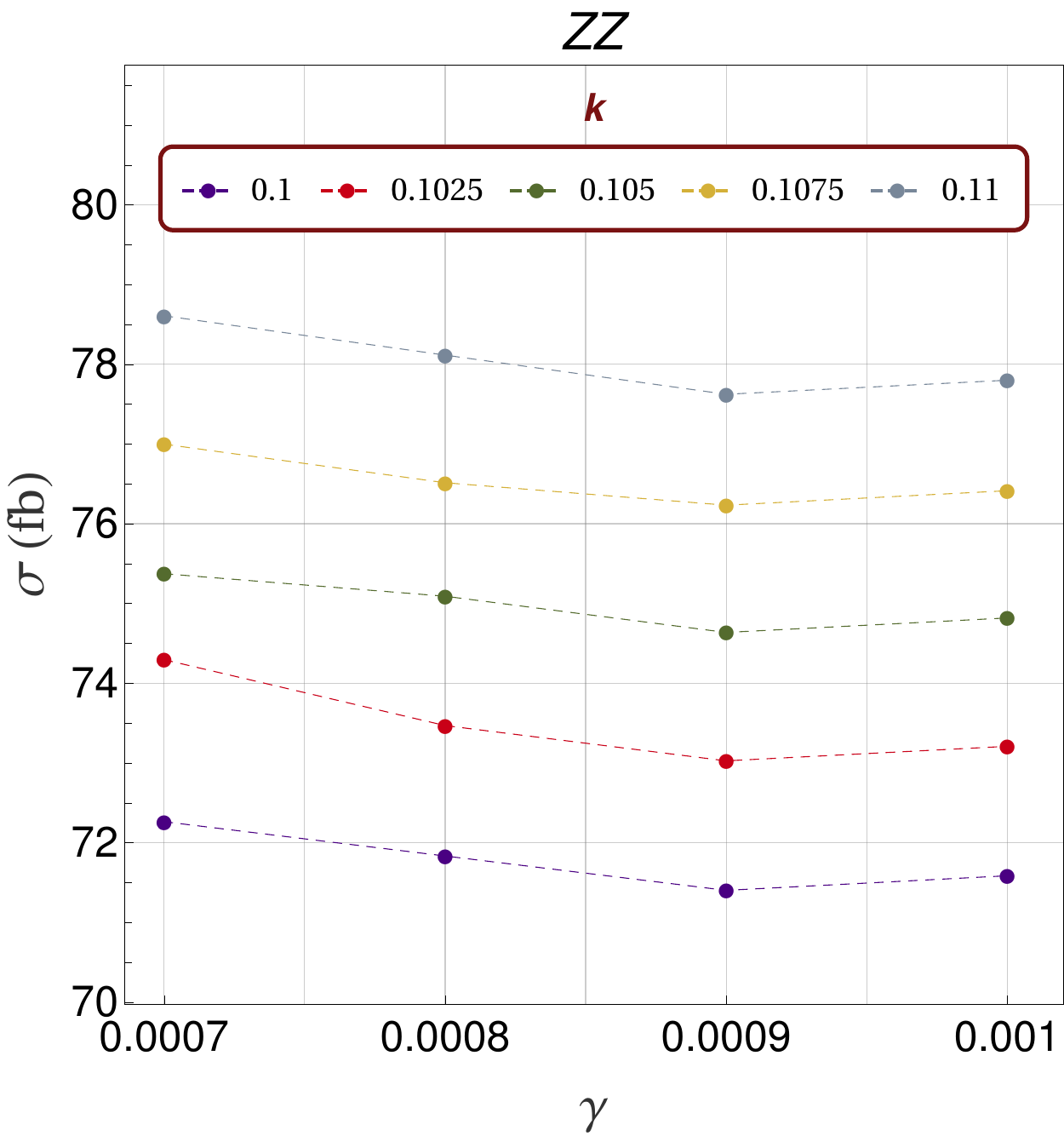}
\caption{\small{Values for the VBS cross section in Eq.~(\ref{eq: total_xs}) with $WW$ (left) and $ZZ$ (right) final states versus the NLO chiral parameter $\gamma$ and for different values of $k=5a_4+8a_5$ in the legend. The combination of the values of $k$ and $\gamma$ present in the figure make up the top-right region of Fig. \ref{fig: map_Kgamma} where resonances close to 650 GeV appear. The centre of mass energy for this calculation is set to $\sqrt{s}=13$ TeV.}}
\label{fig: xsections}
\end{figure*}

Detailed information regarding the H$(650)$ can be found in Ref.~\cite{Kundu:2022bpy}. In this section we connect these tentative results
with the ones extracted from our analysis.
We will work under the assumption that the resonant profile obtained, if any, is produced by a single resonance, neglecting the possibility of
two overlapping resonances. Besides, we also ignore the decay mode H$(650)\to h(125)h(95)$, being $h(125)$ the Higgs described in the minimal SM.
In other words, we ignore the possible presence of the $h(95)$ that seems less motivated. One way to search for H$(650)$ in vector boson fusion
is via the decay to $b\overline{b}\gamma\gamma$, but there is yet not enough resolution in the experiment to distinguish
between a $b\overline{b}$ pair decayed from a $Z$ or a hypothetical h$(95)$. This is why for this work we assume the
decay mode of H$(650)$ to be exclusively via gauge bosons (however, do keep in mind that other channels contribute in the unitarization procedure).\myspace
In Ref.~\cite{Kundu:2022bpy}, the authors gave a total width to gauge bosons of $\Gamma=90\pm 28$ GeV, of the order of the widths presented in the right panel of Fig. \ref{fig: map_Kgamma}.\myspace 
To get a first estimate of the cross section for the production of such resonance we will use the effective $W$ approximation
(EWA)~\cite{Dawson:1984gx}, which takes $W$s and $Z$s as proton constituents and it is approximately valid for energies well above the
EW scale. Within the EWA approach the differential cross-section is given by 
\begin{widetext}
\begin{equation}\label{eq: dxsdMww}
\frac{d\sigma}{dM_{WW}^2}=\sum_{i,j}\int_{M_{WW}^2/s_{}}^{1}\int_{M_{WW}^2/(x_1s_{})}^{1}\frac{dx_1dx_2}{x_1x_2s_{}}f_i(x_1,\mu_F)f_j(x_2,\mu_F)\frac{dL_{WW}}{d\tau}\int_{-1}^{1}\frac{d\sigma_{WW}}{d\cos\theta}d\cos\theta
\end{equation}
\end{widetext}
with $s_{}$ is the centre of mass energy of the two opposite protons at the LHC and $M_{WW}$ is the invariant mass of the two $W$s.
Here the "partonic" differential cross section in the $WW$ rest frame is
\begin{equation}\label{eq: dxsdcos}
\frac{d\sigma_{WW}}{d\cos\theta}=\frac{|A(M_{WW}^2,\cos\theta)|^2}{32\pi M_{WW}^2},
\end{equation}
This expression factorizes both energy scales: the one for the long-distance non-perturbative part describing the dynamics
inside the proton and a perturbative one for the $WW$ hard scattering. \myspace
The amplitude $A(M_{WW}^2,\cos\theta)$ appearing in Eq.~(\ref{eq: dxsdMww}) describes the amplitude of a $WW$ scattering in the charged (physical) basis that is detected at the LHC and not the amplitude in the $IJ$ basis that we rendered unitary. Thus, by moving backwards along the  process we used
for unitarization, we have to recover the "unitary" physical amplitude that would produce such a unitary partial wave. From now on, a superindex $\cal U$ will refer to unitarized quantities, in our case obtained thorough IAM. This is done by reversing the unitarization procedure in Eq.~(\ref{eq: tIAM})
\begin{equation}\label{eq: TIIAM}
T^{\cal U}_{I}=32\pi\sum_{J=0}^{\infty}(2J+1)\,t^{\cal U}_{IJ}P_{J}(\cos\theta)
\end{equation}
and we truncate the infinite series at the leading order (LO) for every isospin channel assuming that is a good approximation close to the
resonance mass where the peak dominates the amplitude:
\begin{equation}\label{eq: TsIAM}
\begin{split}
&T_0^{\cal U}\approx 32\pi\, t^{\cal U}_{00}\\
&T_1^{\cal U}\approx 32\pi\left(3t_{11}^{\cal U}\cos\theta\right)\\
&T_2^{\cal U}\approx 32\pi t_{20}^{\cal U}.
\end{split}
\end{equation}
Now, we simply use the isospin relation to build the "unitary" physical amplitudes. We will be using for comparison
with the literature those with $WW$ and $ZZ$ in the final states after VBS:
\begin{equation}\label{eq: Aphysical}
\begin{split}
A^{\cal U}\left(\wpwm \to \wpwm\right)&=\frac{1}{3}T_0^{\cal U}+\frac{1}{2}T_1^{\cal U}+\frac{1}{6}T_2^{\cal U}\\
A^{\cal U}\left(\wpwm \to ZZ\right)&=\frac{1}{3}T_0^{\cal U}-\frac{1}{3}T_2^{\cal U}\\
A^{\cal U}\left(ZZ\to ZZ\right)&=\frac{1}{3}T_0^{\cal U}+\frac{2}{3}T_2^{\cal U}
\end{split}
\end{equation}
and the symmetric processes under time reversion.\myspace
The EWA consists in convoluting the probability for a quark inside a proton to radiate a gauge boson with the
actual parton distribution function (pdf) for the constituent quarks $q$ at some energy scale, $f_q(x,\mu)$, using the effective luminosity
\begin{equation}\label{eq: luminosity_WW}
\!\!\!\!\frac{dL_{W W}}{d\tau}=\left(\frac{g}{4\pi}\right)^4\left[\left(\frac{1}{\tau}+1\right)\ln\left(\frac{1}{\tau}\right)-2\left(\frac{1}{\tau}-1\right)\right]
\end{equation}
where $\tau=M_{WW}^2/(x_1x_2s_{})$ connects both energy scales. A factor $1/2$ must be added for $ZZ$ final states accounting for their
indistinguishability.\myspace
After performing the convolution of these functions we are in disposition to compute the integral in Eq.~(\ref{eq: dxsdMww}) to obtain the
differential cross section of the process with respect to the invariant mass of the $WW$ system. Moreover, the total cross section of the process
is obtained assuming that the peak indeed dominates the amplitude in such a way that
\begin{equation}\label{eq: total_xs}
\sigma=\int_{M-2\Gamma}^{M+2\Gamma}dM_{WW}\,\frac{d\sigma}{dM_{WW}}
\end{equation}
where $M$ and $\Gamma$ are the characteristic parameters of the resonance obtained from the unitarized amplitudes.\myspace
We are now in disposition to compare our results from the unitarized analysis of $\sigma$ next to the experimental ones. But first, two aspects
need to be taken into account. Firstly, our analysis  only says something about longitudinally polarized gauge bosons in the external states; any
contribution coming from different polarization combinations is to be computed separately and with the corresponding effective luminosity.
However, we expect the purely longitudinal process to dominate at high energies when we separate, even just a little, from the SM. If that
is the case we should not saturate the experimental value which is unpolarized. Secondly, we can easily include in our calculation kinematical
cuts on the pseudo-rapidity and the invariant mass of the outgoing gauge bosons but we can not demand restrictions in the kinematics
of the radiated light jets suitable for VBS detection that are usually included in experimental analysis.\myspace
Taking into account all the machinery developed in this section to compute a theoretical cross section in $pp$ collisions and the comments in the above paragraph, we can now compare with experimental results. In Fig. \ref{fig: xsections} we show values of the crossed sections obtained using Eq.~(\ref{eq: total_xs}) for a subset of the parameter space in Fig. \ref{fig: map_Kgamma} ($k\times\gamma= \left[0.1,0.11\right]\times\left[0.0007,0.001\right]$) where resonances with masses close to 650 GeV appear.\myspace
\myspace
In Fig. \ref{fig: xsections}, we show the values of the cross sections obtained for $WW$ (left panel) and $ZZ$ (right panel) final states at $\sqrt{s}=13$ TeV $pp$ collision energy after VBS using Eq.~(\ref{eq: Aphysical}). The cross sections for WW result to be of order $\sim300$ fb and $\sim$75 fb for $ZZ$ for all values of $k$ and $\gamma$. The measured cross sections from VBS~\cite{Kundu:2022bpy} of $\text{H(650)}\to WW$ and $\text{H(650)}\to ZZ$ are $160\pm 50\,\text{fb}$ and $30\pm 15\, \text{fb}$, respectively, close to SM values. These measurements really favor a $WW$ scenario after VBF rather than a $ZZ$ final state, with a ratio between cross sections $\sigma_{WW}/\sigma_{ZZ}\sim 5$. Our calculation implies $\sigma_{WW}/\sigma_{WW}\sim 4$, relatively close to the ATLAS and CMS analysis but really distant from the prediction using the Georgi-Machacek model, which infers the inverse situation with a $ZZ$ final state dominating over the $WW$ one with $\sigma_{WW}/\sigma_{ZZ}\sim 0.5$~\cite{Georgi:1985nv,Kundu:2022bpy}.   \myspace
We obtain with our calculation, thus, two times the measured central values. As explained before, no cuts, and actually no kinematical condition, are imposed in our calculation in Fig. \ref{fig: xsections} for a better comparison with the experiment so we would expect our computed cross section to exceed the measured one. We can easily introduce a cut in the pseudo-rapitidies of the final state gauge bosons that favors identification of VBS events. In particular if we impose $|\eta_W|<2$~\cite{Delgado:2017cls} the crossed sections are reduced to $\sim 275$ fb and $\sim$70 fb for $WW$ and $ZZ$ processes, respectively, getting closer to the experimental data. Presumably, further cuts on the kinematical variables of the light jets produced after radiation of the gauge bosons triggering the VBS would point towards even closer cross sections.\myspace
To conclude this section we present a test of our calculation by making a comparison in the number of events obtained for a process using MonteCarlo (MC) techniques in Ref.~\cite{Delgado:2017cls}. In this work the authors reproduced the signal expected at the LHC for vector charged resonances emerging in the subprocess $WZ\to WZ$. The range of chiral parameters used lead to resonances in the mass range 1.5-2.5 TeV. For the event simulation at centre of mass energy of $\sqrt{s}=14$ TeV, the authors used a series of kinematical cuts both in the $WZ$ bound state and in the radiated light jets. On one hand, we also introduce for a more reliable comparison the cut on the pseudo-rapidity $|\eta_W|<2$ by integrating the partonic amplitude in the corresponding values of $\cos\theta$. On the other hand, we can not apply any cut on the light jets easily. \myspace 
The number of events obtained from our calculation of the VBS cross section for a specific value of the integrated luminosity is $N_{\mathcal{L}}=\sigma\cdot\mathcal{L}$, where $\sigma$ is the total cross section in Eq.~(\ref{eq: total_xs}). The results for both number of events, the MC simulation in Ref.~\cite{Delgado:2017cls} and our theoretical prediction, are gathered in Table \ref{table: events}:\myspace
\begin{table}[h!]
\centering
\begin{tabular}{|c|c|c|c|c|c|}
\hline
 & $M_V\ih \Gamma_V$ & $N^{\text{MC}}_{1000}$ & $N_{1000} $ & $N^{\text{MC}}_{3000}$ & $N_{3000} $\\ \hline
$ \text{BP1}$ &$1510\ih 13 $ & $488 $ & $904 $ & $1465 $ & $2713 $ \\ [0.6ex]\hline
$ \text{BP2}$ &$2092\ih 20 $ & $82 $ & $121 $ & $246 $ & $364 $ \\ [0.6ex]\hline
$ \text{BP3}$ & $2541\ih 27 $ & $30 $ & $31 $ & $89 $ & $94 $ \\ [0.6ex]\hline
\end{tabular}
\caption{{\small Number of MC events for three benchmark points BP1, P2 and BP3 in Ref.~\cite{Delgado:2017cls} that produce vector resonances for different values of the integrated luminosity $\mathcal{L}$ in units of fb$^{-1}$, $N^{MC}_{\mathcal{L}}$. The number of events obtained with our calculation is $N_{\mathcal{L}}=\sigma\cdot\mathcal{L}$.}}\label{table: events}
\end{table}
As expected and argued before, all our number of events exceed the ones obtained using a MC simulation due to the lack of extra kinematical cuts. We also observe that the heavier the vector resonance is the more accurate the calculation with respect the MC. 
\section{Conclusions}
Searches for light scalar states should be pursued since their existence could be understood as Higgs companions extending the SM and giving an explanation to its origin. Up to now, there are no clear discoveries but there is still room for this new physics to emerge in the high-luminosity phase of the LHC.\myspace
In this work we have seen that H(650) can be accommodated in the HEFT but it requires the cooperation of at least one more next-to-leading
coupling (for which no relevant bounds exist) when the coefficients $a_4$ and $a_5$ are pushed to the limit of the experimentally allowed region.
Further restrictions on them derived from experiment would probably hinder the viability of the tentative resonance H(650). The prediction for the
width of this resonance also fits well with the preliminary experimental observations.\myspace
We computed within the EWA, the cross section for the production of a 650 GeV resonance in the vector boson fusion channel via $pp\to WW+X$ and compared it with the results in Ref.~\cite{Kundu:2022bpy}, assuming that the scalar state only decays in gauge bosons and not to any other two scalars. We also include a comparison between the number of events with a MC simulating a charged vector resonance in the process $WZ\to WZ$~\cite{Delgado:2017cls}.\myspace
The results we find are encouraging in the sense that the predicted cross section is relatively close to the experimental analyses
performed by ATLAS and CMS. First, without taking into account any event selection cut we obtained cross sections of $\sim 300$ fb and $\sim 75$ fb
for $WW$ and $ZZ$ final states, respectively. These results are obtained within the EWA and assuming that the peak dominates the cross section. After applying cuts on the pseudorapidty of the di-boson state, the values of the cross sections are reduced to $\sim 275$ fb and $\sim 70$ fb for $WW$ and $ZZ$ states, respectively.
%%%%%%%%%%%%%%%%

\begin{small}

\bibliographystyle{utphys}
\bibliography{bibliography}
 \end{small}

\end{document}